\documentclass{Interspeech}
\usepackage{cite}
\interspeechcameraready

\title{STSR: High-Fidelity Speech Super-Resolution via Spectral-Transient Context Modeling}

\author[affiliation={1}]{Jiajun}{Yuan}
\author[affiliation={1}]{Xiaochen}{Wang}
\author[affiliation={1}]{Yuhang}{Xiao}
\author[affiliation={2}]{Yulin}{Wu}
\author[affiliation={3}]{Chenhao}{Hu}
\author[affiliation={3}]{Xueyang}{Lv}

\affiliation{National Engineering Research Center for Multimedia Software, School of Computer Science}{Wuhan University, Wuhan}{China}
\affiliation{School of Artificial Intelligence}{Jianghan University, Wuhan}{China}
\affiliation{}{Xiaomi Corporation, Beijing}{China}
\email{yjj2002@whu.edu.cn, wuyulin@whu.edu.cn, huchenhao@xiaomi.com}


\graphicspath{{./}{../swinGAN/}}

\begin{document}

\maketitle

\begin{abstract}
Speech super-resolution (SR) reconstructs high-fidelity wideband speech from low-resolution inputs—a task that necessitates reconciling global harmonic coherence with local transient sharpness. While diffusion-based generative models yield impressive fidelity, their practical deployment is often stymied by prohibitive computational demands. Conversely, efficient time-domain architectures lack the explicit frequency representations essential for capturing long-range spectral dependencies and ensuring precise harmonic alignment. We introduce STSR, a unified end-to-end framework formulated in the MDCT domain to circumvent these limitations. STSR employs a Spectral-Contextual Attention mechanism that harnesses hierarchical windowing to adaptively aggregate non-local spectral context, enabling consistent harmonic reconstruction up to 48 kHz. Concurrently, a sparse-aware regularization strategy is employed to mitigate the suppression of transient components inherent in compressed spectral representations. STSR consistently outperforms state-of-the-art baselines in both perceptual fidelity and zero-shot generalization, providing a robust, real-time paradigm for high-quality speech restoration.
\ifinterspeechfinal
Samples can be found online\footnote{\url{https://judgeyjj.github.io/}}.
\else
[to ensure author anonymity, the link to the samples will be added after the review process]
\fi
\end{abstract}
\begin{figure*}[t]
    \centering
    \includegraphics[width=\linewidth]{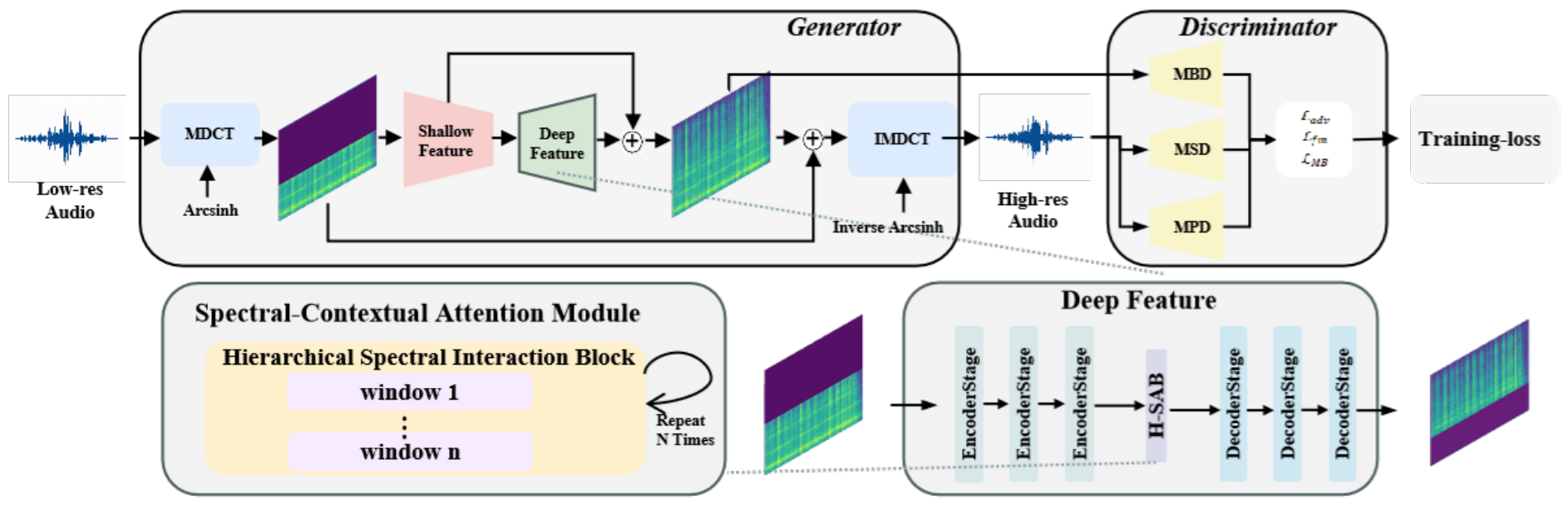}
\caption{Overall architecture of STSR.}
    \label{fig:model_overview}
\end{figure*}
\section{Introduction}

Speech super-resolution (SR), or bandwidth extension (BWE), entails reconstructing high-fidelity wideband spectra from bandwidth-limited inputs. This task remains pivotal for augmenting intelligibility and perceptual quality in real-time telecommunications and the restoration of legacy recordings~\cite{liu2022voicefixer, lu2024highqualityefficientspeechbandwidth}.

Prior discriminative deep learning models~\cite{kuleshov2017aecnn, liu2018tfnet} primarily formulate SR as a deterministic regression task. While these methods successfully recover low-frequency envelopes, their reliance on minimizing sample-wise L1 or L2 distances inevitably precipitates a regression-to-the-mean effect. This yields over-smoothed high-frequency spectra that lack acoustic crispness---a deficiency that becomes particularly pronounced as target sampling rates reach 48~kHz.

To synthesize more realistic spectral textures, recent research has embraced generative modeling. Advanced frameworks such as VoiceFixer, WSRGlow, and AudioSR~\cite{liu2022voicefixer, kim2021wsrglow, liu2023audiosr, moliner2024babe, yun2025flowhighefficienthighqualityaudio} capitalize on iterative priors to hallucinate fine-grained high-band details. Although these diffusion-style or flow-based refinement processes significantly enhance perceptual fidelity, the necessity of multi-step sampling imposes prohibitive computational overhead and latency, rendering them impractical for real-time deployment.

Driven by the need for efficiency, recent efforts have shifted toward end-to-end architectures to maximize computational throughput and circumvent error accumulation. Representative models, such as the time-domain state-space framework Wave-U-Mamba~\cite{lee2025waveumambaendtoendframeworkhighquality, yu2024baenet, liu2025hwbnet, liu2025neuralcodec, lu2024multistagespeechbandwidthextension, tamiti2025highfidelityspeechsuperresolution} and the unified Transformer--convolution architecture HiFi-SR~\cite{zhao2025hifisr}, offer a distinct advantage: they enable rapid inference and eliminate the feature mismatch inherent in two-stage pipelines like NVSR~\cite{liu2022nvsr}. However, while operating exclusively in the waveform domain effectively captures temporal evolution, it lacks an explicit representation of frequency structures. Consequently, time-domain modeling alone struggles to capture the non-local vertical correlations between low-frequency fundamentals and their distant high-band harmonics, a limitation that significantly constrains reconstruction fidelity at 48~kHz.

An alternative paradigm performs vocoder-free SR directly in the Modified Discrete Cosine Transform (MDCT) domain. By bypassing magnitude-to-waveform conversion, this approach circumvents the artifacts typical of two-stage pipelines. Despite these structural merits, current MDCT-based solutions like mdctGAN~\cite{gantt2022mdctgan, bao2025frequency} remain suboptimal. Their reliance on standard convolutional kernels restricts modeling to local spectral neighborhoods, hindering the maintenance of harmonic consistency across the wideband spectrum. Moreover, the heavy-tailed nature of MDCT coefficients necessitates aggressive range compression (e.g., arcsinh) for training stability; unfortunately, this preprocessing tends to flatten sharp spectral peaks, leading to the over-smoothing of transient components and a loss of perceptual attack.

In this paper, we present STSR, a unified MDCT-domain framework designed to reconcile long-range spectral consistency with transient fidelity while meeting the stringent requirements of real-time speech SR. We treat the spectrogram as a structured representation and employ a hierarchical window attention mechanism to adaptively aggregate non-local spectral context, ensuring precise harmonic alignment across the band. To offset the information loss induced by spectral compression, we introduce a sparse-aware loss that enforces transient constraints and preserves acoustic sharpness. Finally, a high-frequency-focused hybrid discriminator concentrates adversarial pressure on synthesized spectra while maintaining global temporal coherence, enabling real-time single-channel upsampling with robust zero-shot generalization.

Our main contributions are as follows:
\begin{enumerate}
    \item A spectral--transient aware MDCT framework that jointly addresses envelope reconstruction and transient preservation;
    \item A spectral-context attention mechanism that explicitly captures non-local dependencies along the frequency axis;
    \item A high-band-focused hybrid discriminator that concentrates adversarial pressure on generated frequencies without destabilizing the low-frequency base;
    \item A sparse-aware loss tailored to compressed MDCT representations, significantly mitigating transient over-smoothing.
\end{enumerate}

\section{Method}
\label{sec:method}
The STSR framework operates directly within the signed MDCT domain. Unlike magnitude-only spectrograms, MDCT coefficients implicitly encapsulate both energy and phase information via their signed values. To accommodate negative values where standard logarithmic compression is ill-defined, we employ a pseudo-logarithmic mapping strategy commonly utilized in high-fidelity audio synthesis~\cite{gantt2022mdctgan}. Specifically, we leverage the inverse hyperbolic sine ($\operatorname{arcsinh}$), which approximates logarithmic scaling for high-amplitude components while maintaining sign integrity and linearity near the origin. Operationally, the MDCT coefficients $X$ are scaled by a gain factor $g$ (where $g{=}800$) and transformed via base-10 pseudo-logarithmic compression:
\begin{equation}
\label{eq:compression}
\begin{aligned}
    S &= \frac{\operatorname{arcsinh}(g \cdot X)}{\ln(10)}, \\
    X &= \frac{\sinh(S \cdot \ln(10))}{g}.
\end{aligned}
\end{equation}
This transformation effectively regularizes the heavy-tailed distribution of spectral data while ensuring perfect invertibility, thereby facilitating robust waveform reconstruction.

\subsection{Spectral Branch (Generator)}
\label{ssec:generator}
Although $\operatorname{arcsinh}$ compression facilitates training stability, it inherently attenuates fine-grained spectral details and transient cues. Standard convolutional generators often fail to recover these degraded components, as their representation power is fundamentally constrained by the locality of their receptive fields. Drawing upon recent advances in attention-based restoration~\cite{liu2021swin, liang2021swinir, choi2022swin2sr}, we propose a Hierarchical Spectral-Contextual Attention mechanism explicitly tailored to the spectral anisotropy of speech.

Since global self-attention across the entire spectrogram is precluded by prohibitive computational complexity, we implement Spectral Interaction Blocks that leverage windowed self-attention. This design partitions the frequency axis into local windows to capture localized harmonic structures, including fundamental frequencies and their immediate partials. To facilitate information flow across these boundaries, a shifted-window mechanism is employed in alternating layers. This strategy enables the modeling of long-range spectro-temporal evolutions---such as formant transitions---that are essential for speech SR yet remain elusive for conventional CNNs.

We architect a compact U-Net using these attention blocks, incorporating skip connections to preserve low-level spectral features. Additionally, a global residual path is integrated to bias the model toward explicitly learning high-frequency residuals. The resulting architecture constitutes a unified, single-stage vocoder-free framework that effectively reconciles local transient sharpness with global harmonic consistency.

\subsection{Hybrid Discriminator}
\label{ssec:discriminator}
The discriminator architecture is pivotal in synthesizing high-fidelity waveforms. We leverage the Multi-Period (MPD) and Multi-Scale (MSD) discriminators from HiFi-GAN~\cite{kong2020hifi} to enforce temporal coherence and suppress phase artifacts. While these time-domain modules effectively capture periodic structures and global patterns, they exhibit limited sensitivity to fine-grained spectral artifacts within the reconstructed high-frequency regime. 

To rectify this, we introduce a High-Band Multi-Band Discriminator (HB-MBD) that operates directly on MDCT magnitudes. Given that the low-frequency band is provided as ground-truth input, subjecting this region to adversarial pressure can induce spectral degradation or optimization instability. We therefore delineate frequency boundaries based on sampling rates to isolate the reconstruction target: let $f_{hi}$ and $f_{lo}$ denote the Nyquist frequencies of the target high-resolution output and the low-resolution input, respectively. 

To preserve the integrity of the low-band, discriminative focus is restricted to the interval $[f_{lo}, f_{hi})$. Specifically, parallel PatchGAN heads are deployed across distinct sub-bands uniformly distributed within this range, complemented by a full-band head to ensure global spectral consistency. This configuration confines adversarial feedback to the synthesized components, preventing the model from compromising low-band fidelity for high-band hallucination. The resulting optimization yields crisper harmonics and fricatives without destabilizing the fundamental frequencies.

\begin{figure*}[t]
    \centering
    \includegraphics[width=\linewidth]{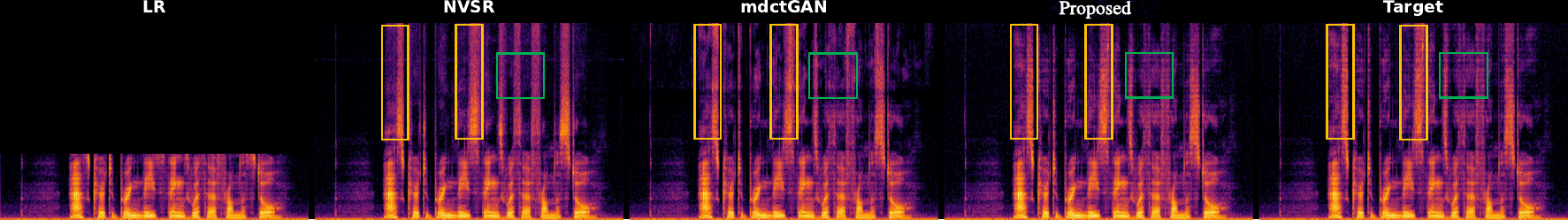}
    \caption{Spectrogram comparison (16kHz$\rightarrow$48kHz). STSR exhibits sharper harmonic structures, more complete high-band energy, and clearer transients with fewer artifacts, surpassing prior models in both detail and fidelity.}
    \label{fig:spec}
    \vspace{-0.3cm}
\end{figure*}

\subsection{Training Objectives}
\label{ssec:training_loss}
STSR is optimized via a composite objective function designed to reconcile the trade-off between perceptual fidelity and spectral precision.

\textbf{Adversarial Loss.} To synthesize realistic waveforms, we leverage the Least Squares GAN (LS-GAN) formulation~\cite{mao2017least}. The discriminator ensemble $\mathcal{D}$ comprises time-domain modules (MPD and MSD) and the proposed frequency-domain HB-MBD, denoted as $D_{wav}$ and $D_{spec}$, respectively. Given the ground-truth audio $y$ and the low-resolution input $x$, the generator $G$ and discriminators $D$ engage in the following minimax game:
\begin{equation}
\begin{aligned}
    \mathcal{L}_{adv}(D) &= \mathbb{E}_{y}[(1 - D(y))^2] + \mathbb{E}_{x}[D(G(x))^2] \\
    \mathcal{L}_{adv}(G) &= \mathbb{E}_{x}[(1 - D(G(x)))^2]
\end{aligned}
\end{equation}
The total adversarial objective is formulated as a weighted sum across all discriminator sub-units in both the time and frequency domains.

\textbf{Feature Matching Loss.} We incorporate feature matching loss to stabilize the adversarial landscape and facilitate high-fidelity synthesis~\cite{kong2020hifi}. This loss minimizes the $L_1$ distance between internal feature maps of the discriminators for real and synthesized samples:
\begin{equation}
    \mathcal{L}_{fm}(G) = \sum_{k=1}^{K} \sum_{l=1}^{L_k} \frac{1}{N_l} \left\lVert D_k^{(l)}(y) - D_k^{(l)}(G(x)) \right\rVert_1
\end{equation}
where $K$ denotes the number of discriminators, $L_k$ is the layer count for the $k$-th discriminator, and $N_l$ is the number of elements within the $l$-th feature map.

\textbf{Transient Constraint (Sparse-Aware Loss).} While standard $L_1$ reconstruction losses treat spectral bins uniformly, MDCT coefficients are inherently sparse, characterized by high-energy harmonics and transient peaks set against a near-zero background. Although $\operatorname{arcsinh}$ compression is essential for numerical stability, it inevitably compresses the dynamic range, which may amplify background artifacts and blur transient boundaries. A naive global $L_1$ objective fails to differentiate these regimes, often yielding over-smoothed reconstructions where transient attacks lose their acoustic crispness.

To rectify this, we propose a content-adaptive sparse loss that decouples signal reconstruction from noise suppression. We formulate a dynamic soft mask derived from the ground-truth energy profile. Let $S$ and $\hat{S}$ represent the compressed signed MDCT coefficients of the ground truth $y$ and the prediction $G(x)$, respectively. A content weight map $w_c$ is computed via a sigmoid function centered on the spectral envelope:
\begin{equation}
\begin{aligned}
    w_c &= \sigma(\alpha \cdot (|S| - \tau)) \\
    w_s &= 1 - w_c
\end{aligned}
\end{equation}
where $\sigma$ denotes the sigmoid function and $\alpha$ modulates the transition steepness. Rather than utilizing a fixed threshold, $\tau$ is dynamically determined as the 0.8-quantile of $|S|$ for each utterance, ensuring robust performance across diverse volume levels and recording environments.

The resulting objective combines two complementary terms:
\begin{equation}
    \mathcal{L}_{sparse}(G) = \underbrace{\left\lVert w_c \odot (S - \hat{S}) \right\rVert_1}_{\text{Signal Fidelity}} + \lambda_{s} \underbrace{\left\lVert w_s \odot \hat{S} \right\rVert_1}_{\text{Sparsity Regularization}}
\end{equation}
The former prioritizes active spectral regions (harmonics and transients), enforcing fidelity where energy is prominent. The latter penalizes spurious activations in background regions, effectively suppressing GAN-induced artifacts in sparse frequency bins. By re-weighting the optimization landscape, this strategy counteracts compression-induced smoothing and restores the natural sparsity of the speech signal.

\textbf{Total Objective.} The global loss function integrates these terms with a multi-resolution STFT loss $\mathcal{L}_{stft}$~\cite{yamamoto2020parallel} to ensure stable convergence:
\begin{equation}
\begin{aligned}
    \mathcal{L}_{G} &= \lambda_{adv}^{(t)} (\mathcal{L}_{adv}^{wav} + \mathcal{L}_{adv}^{spec}) + \lambda_{fm}\mathcal{L}_{fm} \\
    &\quad + \lambda_{stft}\mathcal{L}_{stft} + \lambda_{sparse}\mathcal{L}_{sparse}
\end{aligned}
\end{equation}
To maintain optimization stability in the MDCT domain, a linear warm-up schedule is applied to the adversarial weight $\lambda_{adv}^{(t)}$, allowing the generator to capture fundamental spectral structures prior to full adversarial refinement.

\section{Experiments}
\label{sec:experiments}

\subsection{Datasets}
\label{ssec:dataset}
STSR was trained and evaluated on the VCTK corpus~\cite{veaux2019vctk}, which encompasses approximately 44 hours of 48~kHz recordings from 110 English speakers. To ensure data integrity, we utilized the \texttt{mic1} channel and excluded speakers \texttt{p280} and \texttt{p315} following established protocols. The remaining data were partitioned into 100 speakers for training and 8 for evaluation. To assess cross-domain robustness, we performed zero-shot evaluation on the HiFi-TTS dataset~\cite{bakhturina2021hi}, which contains high-fidelity 44.1~kHz recordings. To maintain pipeline consistency, these recordings were resampled to 48~kHz prior to processing.

\subsection{Evaluation Metrics}
\label{ssec:eval_metric}

Model performance was quantitatively assessed using Log-Spectral Distance (LSD)~\cite{gray1976distance}. Let \(S\) and \(\tilde{S}\) denote the magnitude spectrograms of the ground-truth and synthesized speech signals, respectively. The LSD is defined as:
\begin{equation}
\begin{aligned}
\text{LSD}(S, \tilde{S}) &= \frac{1}{T} \sum_{t=1}^{T} \sqrt{ \frac{1}{F} \sum_{f=1}^{F} \left( \log_{10} \frac{S(t,f)^2}{\tilde{S}(t,f)^2} \right)^2 }
\end{aligned}
\label{eq:lsd_simplified}
\end{equation}
where $T$ and $F$ represent the number of time frames and frequency bins, respectively. As a frequency-domain metric, LSD quantifies the logarithmic divergence between spectra in decibels (dB). Lower LSD values reflect superior perceptual fidelity, with a value of zero signifying identical spectral profiles. Comparative results against baseline systems, including WSRGlow~\cite{kim2021wsrglow}, NVSR~\cite{liu2022nvsr}, Wave-U-Mamba~\cite{lee2025waveumambaendtoendframeworkhighquality}, and mdctGAN~\cite{gantt2022mdctgan}, are summarized in Table~\ref{tab:lsd_results_compact}.

\begin{figure}[t]
    \centering
    \includegraphics[width=\linewidth]{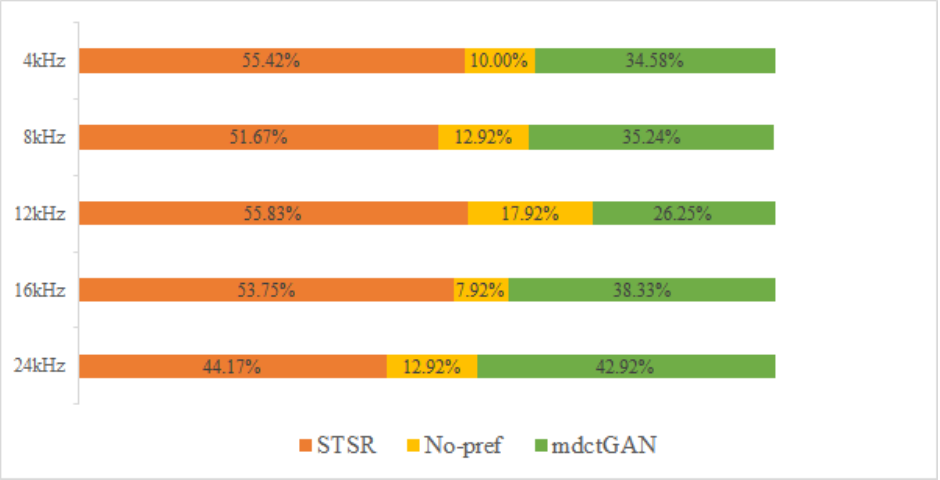}
    \vspace{-0.3cm}
    \caption{ABX subjective test results of mdctGAN and STSR}
    \label{fig:abx}
    \vspace{-0.3cm}
\end{figure}

For subjective evaluation, we conducted an ABX listening test. In each trial, listeners were presented with two system outputs (A and B) alongside a reference (X) and tasked with identifying which sample exhibited higher similarity to the reference. This streamlined protocol facilitates a direct comparison of perceptual quality while minimizing participant fatigue; aggregated preference scores are reported in Figure~\ref{fig:abx}.
 \begin{table}[t]
  \caption{Objective evaluation results (LSD) and model size (Params, M). Lower LSD is better (The best results have been bolded).\\
  }
  \label{tab:lsd_results_compact}
  \centering
  { 
  \setlength{\tabcolsep}{3pt} 
  \footnotesize
  \vspace{-0.3cm}
  \begin{tabular}{lcccccc} 
    \toprule
    \textbf{Model} & \textbf{4kHz} & \textbf{8kHz} & \textbf{16kHz} & \textbf{24kHz} & \textbf{AVG} & \textbf{Params} \\
    \midrule
    Unprocessed      & 6.08 & 5.15 & 4.85 & 3.84 & 4.98 & - \\
    WSRGlow & 1.12 & 0.98 & 0.85 & 0.79 & 0.94 & 229.9M$\times$4 \\
    NVSR & 0.98 & 0.91 & 0.81 & 0.70 & 0.85 & 99.0M \\
    Wave-U-Mamba & 0.97 & 0.87 & 0.75 & 0.70 & 0.82 & \textbf{4.2M} \\
    mdctGAN	& 1.09 & 0.93 & 0.83 & 0.71 & 0.89 & 101.0M \\
    \midrule
    \textbf{STSR} & \textbf{0.95} & \textbf{0.84} & \textbf{0.73} & \textbf{0.65} & \textbf{0.79} & 66.2M \\
    (w/o HB-MBD) & 0.99 & 0.86 & 0.77 & 0.71 & 0.83 & - \\
    (w/o Sparse) & 0.98 & 0.85 & 0.74 & 0.69 & 0.81 & - \\
    \bottomrule
  \end{tabular}
  }
  \vspace{-0.3cm}
 \end{table}

\subsection{Implementation Details}
\label{ssec:training}
STSR was trained on a single NVIDIA A100 GPU for 200k iterations with a batch size of 16. To mitigate overfitting to fixed spectral cutoffs, low-resolution inputs were synthesized on-the-fly: during each iteration, a cutoff frequency $r \in [4, 32]$~kHz was stochastically sampled, followed by low-pass filtering, downsampling, and subsequent upsampling back to 48~kHz. This dynamic degradation strategy compels the model to reconstruct diverse spectral bands from 48,460-sample audio segments.

The framework utilizes a four-stage hierarchical encoder-decoder architecture optimized to balance model capacity with computational efficiency. Encoder stages were configured with depths of $\{2, 2, 8, 2\}$, while the number of attention heads scaled from 4 to 32 to facilitate robust multi-scale dependency modeling. Input MDCT coefficients were computed using a 1024-sample window and a 512-sample hop length, utilizing a Kaiser-Bessel Derived (KBD) window ($\alpha{=}6$) with an $\operatorname{arcsinh}$ gain of $g{=}800$. Optimization was conducted via the AdamW optimizer ($\beta_1{=}0.8$, $\beta_2{=}0.99$, weight decay $0.01$) with an initial learning rate of $2\times10^{-4}$ and an exponential decay factor of $0.999$ per epoch. Adversarial weights $\lambda_{wav}^{(t)}$ and $\lambda_{spec}^{(t)}$ underwent a linear warm-up over the initial 20k iterations to stabilize the nascent training phase.
\begin{table}[t]
    \caption{Cross-dataset generalization on HiFi-TTS (LSD). Lower is better. Zero-shot: no fine-tuning (The best results have been bolded).\\
}
\vspace{-0.3cm}
    \label{tab:hifitts_generalization}
   \centering
   { 
   \setlength{\tabcolsep}{3pt} 
   \footnotesize
   \vspace{-0.3cm}
   \begin{tabular}{lccccc}
     \toprule
     \textbf{Model} & \textbf{4kHz} & \textbf{8kHz} & \textbf{16kHz} & \textbf{24kHz} & \textbf{AVG} \\
     \midrule
     NVSR & 1.39 & 1.23 & 0.93 & 0.82 & 1.09 \\
     mdctGAN & 1.41 & \textbf{1.21} & 0.95 & 0.84 & 1.10 \\
     \textbf{STSR} & \textbf{1.33} & 1.21 & \textbf{0.89} & \textbf{0.78} & \textbf{1.05} \\
     \bottomrule
   \end{tabular}
   }
   \vspace{-0.3cm}
 \end{table}

\section{Results and Discussion}
\label{sec:results}
Table~\ref{tab:lsd_results_compact} summarizes objective performance on the VCTK test set. STSR attains an average LSD of 0.79, consistently surpassing the regression-based NVSR (0.85). Compared to the lightweight Wave-U-Mamba~\cite{lee2025waveumambaendtoendframeworkhighquality} (4.2M parameters), STSR yields superior spectral fidelity (0.79 vs. 0.82), with the performance gain most evident in the challenging 16~kHz$\rightarrow$48~kHz task (0.73 vs. 0.75). These results suggest that while time-domain state-space models are parameter-efficient, our MDCT-domain attention mechanism more effectively captures the fine-grained harmonic structures essential for high-fidelity restoration, justifying the associated computational cost (66.2M parameters).

Ablation studies underscore the efficacy of each proposed module (Table~\ref{tab:lsd_results_compact}). Excising the HB-MBD increases the average LSD to 0.83, confirming that frequency-domain adversarial feedback is pivotal for refining high-band textures. Similarly, omitting the Transient Constraint degrades the score to 0.81, validating that sparsity-aware regularization effectively mitigates the smoothing artifacts induced by $\operatorname{arcsinh}$ compression.

To evaluate generalization, VCTK-trained models were tested directly on the unseen HiFi-TTS dataset~\cite{bakhturina2021hi} without fine-tuning. As shown in Table~\ref{tab:hifitts_generalization}, STSR yields a superior average LSD of 1.05, outperforming both NVSR (1.09) and mdctGAN (1.10). This consistent performance across distribution shifts indicates that STSR learns robust, generalized spectro-temporal dependencies.

Subjectively, ABX listening tests (Figure~\ref{fig:abx}) reveal a distinct preference for STSR over the mdctGAN baseline. This preference is most pronounced at 16~kHz input bandwidths—a critical threshold where fundamental frequencies are preserved but high-frequency harmonics are entirely absent—confirming that STSR delivers perceptually crisper audio.

\section{Conclusions}
\label{sec:conclusion}
We presented STSR, a MDCT-domain framework designed to reconcile global harmonic coherence with local transient sharpness. Our architecture integrates a hierarchical spectral-contextual attention mechanism to capture long-range dependencies and a sparse-aware regularization strategy to counteract compression-induced smoothing. Experimental results demonstrate that STSR surpasses state-of-the-art baselines in both objective metrics and perceptual quality, while exhibiting robust zero-shot generalization. Notably, the model achieves a real-time throughput of 12.5$\times$ on a single NVIDIA A100 GPU, satisfying the stringent latency requirements of practical applications without compromising reconstruction fidelity.

\ifinterspeechfinal
\section{Acknowledgements}
This work was supported by the National Natural Science Foundation of China under Grant No. 62271358.
\fi

\bibliographystyle{IEEEtran}
\bibliography{references}

\end{document}